\documentclass[dvips]{article}
\usepackage{icrctc07}

\title{Observations of 1ES 0647+250 and 1ES 0806+524 with VERITAS}
\shorttitle{1ES 0647+250 and 1ES 0806+524 with VERITAS}
\authors{P. Cogan$^1$ for the VERITAS Collaboration$^2$.}
\shortauthors{P. Cogan and et al}
\afiliations{$^1$McGill University, 3600 University Street, Montreal, QC H3A 2T8, Canada \\ 
$^2$For full author list see G. Maier, "Status and Performance of VERITAS", these proceedings}
\email{coganp@hep.physics.mcgill.ca}

\abstract{Observations of the blazars 1ES 0647+250 and 1ES 0806+524
with VERITAS are reported here. These objects are among the favoured
candidate extragalactic sources in the very high-energy regime due to
the presence of high-energy electrons and adequate seed photons. The
presence of high-energy electrons is established from the location of
the synrchrotron peak in the spectral energy distribution of the
blazars. The presence of adequate seed photons is determined by the
flux in the radio-through-optical wavebands. These are the key
ingredients for very high-energy gamma-ray emission in the context of
the synchrotron self-Compton model. The redshift of 1ES 0647+250 has
been tentativeley reported as 0.203 and the redshift of 1ES 0806+524
is 0.138, thus the detection of very high-energy gamma-ray emission
from these objects could make significant contributions to the
understanding of the extragalactic infrared background light. The
analysis of these data relies on standard techniques in very
high-energy gamma-ray astronomy, and the results are compared to
previously reported upper limits and to theoretical predictions.}

\begin{document}
\maketitle

\section{Introduction}

The blazars 1ES 0647+250 and 1ES 0806+524 are among several blazars
observed with VERITAS during its commissioning phase in late 2006 and
early 2007. The observations of 1ES 0647+250 and 1ES 0806+524 are
motivated by the search for very high-energy gamma-ray emission from
extragalactic sources. In this paper two specific models used to
predict VHE emission from blazars will be discussed and previously
reported measurements and flux upper limits of the objects compared. The
VERITAS instrument is briefly described and the results from the
analysis of the VERITAS data are reported.

\section{Predictive Models}

In the \emph{modified-Fossati} model \cite{costamante02,fossati98}, the peak
frequency of the synchrotron spectrum and the relative importance of
the inverse-Compton power are determined by the radio luminosity. The
first modification assumes that objects of low power
have equal luminosities in the synchrotron and self-Compton components
of their spectra. The second modification extends
the radio range below $10^{41.2}\:\mathrm{erg}\:\mathrm{s}^{-1}$ by
modeling Compton scattering in the Klein-Nishina regime. This
modification also uses a different width for the parabola representing
the Compton peak, which is reduced with respect to the synchrotron
peak.

In the \emph{Costamante} \cite{costamante02} model, a single zone SSC
fit to multiwavelength data is used to predict fluxes in the TeV
regime. The model emphasises the requirement of \emph{both}
high-energy electrons and sufficient seed photons to produce very
high-energy gamma rays. In the study, a large sample of BL Lacs were
examined and the radio/optical flux and synchrotron peak frequency fit
using the SSC model. The flux predictions are summarised in Table \ref{table1}.

\begin{table}[htbp]
  \begin{center}
    \begin{tabular}{ |c|c|c|c|}
      \hline
      Object       & \emph{Costamante} & \emph{Modified-Fossati} \\ \hline
      1ES 0647+250 &  0.59             &  0.24 \\ \hline
      1ES 0806+524 &  1.36             &  -    \\ \hline
    \end{tabular}
  \end{center}
  \caption{Flux predictions according to the \emph{Costamante} and
  \emph{modified-Fossati} models. Fluxes are in units of
  $10^{-11}\:\mathrm{cm}^{-2}\:\mathrm{s}^{-1}$ above 0.3 TeV.}
  \label{table1}
\end{table}

\section{Targets}

The blazar 1ES 0647+250 was discovered in the MHT-Green Bank survey at
5 GHz using the NRAO 91-m transit telescope. X-ray
emission was discovered in the Einstein Slew Survey with the
synchrotron peak falling just below 10 keV. A redshift of
0.203 has been tentatively reported. Previous observations of this object
were reported by HEGRA \cite{aharonian04_hegra54} with a $99\:\%$
confidence flux upper limit of
$\mathcal{F}_{\mathrm{E}>0.78\:\mathrm{TeV}}<33.5 \times
10^{-11}\:\mathrm{cm}^{-2}\:\mathrm{s}^{-1}$ from 4.1 hours of
observations, where a spectral index of -2.5 was assumed. To compare
this to the \emph{Costamante} and \emph{modified-Fossati} models, this
can be extrapolated to
$\mathcal{F}_{\mathrm{E}>0.3\:\mathrm{TeV}}<126.4 \times
10^{-11}\:\mathrm{cm}^{-2}\:\mathrm{s}^{-1}$ assuming a constant
spectral index of -2.5 down to 0.3 TeV. This is well above
both the predictions of the \emph{Costamante} and
\emph{modified-Fossati} models and does not constrain either model.

The blazar 1ES 0806+524 was discovered using the NRAO Green Bank 91-m
telescope at 4.85 GHz with X-ray emission reported by the
Einstein Slew Survey. The galaxy has a measured redshift of
0.138. Whipple \cite{horan04_constraints,delacalleperez03} reported
flux upper limits of $\mathcal{F}_{\mathrm{E}>0.3\:\mathrm{TeV}}<1.37
\times 10^{-11}\:\mathrm{cm}^{-2}\:\mathrm{s}^{-1}$,
$\mathcal{F}_{\mathrm{E}>0.3\:\mathrm{TeV}}<16.80 \times
10^{-11}\:\mathrm{cm}^{-2}\:\mathrm{s}^{-1}$ and
$\mathcal{F}_{\mathrm{E}>0.3\:\mathrm{TeV}}<1.47 \times
10^{-11}\:\mathrm{cm}^{-2}\:\mathrm{s}^{-1}$ from different observing
seasons. HEGRA reported an upper limit of
$\mathcal{F}_{\mathrm{E}>1.09\:\mathrm{TeV}}<42.5 \times
10^{-11}\:\mathrm{cm}^{-2}\:\mathrm{s}^{-1}$ in one hour of
observations which can be extrapolated to
$\mathcal{F}_{\mathrm{E}>0.3\:\mathrm{TeV}}<255.4 \times
10^{-11}\:\mathrm{cm}^{-2}\:\mathrm{s}^{-1}$ (assuming a spectral
index of -2.5) to compare with the Whipple result and
\emph{Costamante} prediction. Neither the Whipple nor the HEGRA
results constrain the \emph{Costamante} model.

\section{Observations}

All observations were taken in \textsc{wobble} mode. In this mode, the
target is offset from the center of the field of view by
$\pm0.3^\circ$ or $\pm0.5^\circ$ in declination (or right ascension). One of the
trade-offs with the \textsc{wobble} mode is that although more time
can be spent with the observational target in the field of view, the
target is not at the center of the field of view where the sensitivity
is highest. Thus the \textsc{wobble} offset must be carefully chosen
using measurements or Monte Carlo simulations of the detector response to maximise
sensitivity. All observations were taken during the VERITAS
commissioning phase, where initially two telescopes and later three
telescopes were complete. After selecting data to remove runs
suffering from bad weather or technical problems, a total of 17.3
hours on 1ES 0647+250 and 34.7 hours on 1ES 0806+524 were available. 

\section{Data Analysis}

The data have been analysed using independent analysis packages (see
\cite{cogan07} for details on the analysis package). All of these
analyses yield consistent results. Standard data analysis techniques
\cite{daniel07} for ground-based gamma-ray astronomy were used, and
are briefly described here. The analysis was optimised using Crab
Nebula \cite{celick07} data from the same period.

Shower images in the focal plane are gain corrected and cleaned before
being parameterised using a moment analysis. For each
shower, the focal plane images are parameterised using Mean-Scaled
Width and Length (MSW/MSL)\cite{daniel07}. These are calculated using
a large data set of Monte Carlo simulations at discrete zenith
($\Theta$) angles (with interpolation in $\cos \Theta$). Cuts on MSL
and MSW are designed to reject most of the background while retaining
a large portion of the signal. The cuts for this analysis were
optimised on a data set of Crab Nebula observations. 

Background estimation is performed using the reflected-region
background model. In this scheme, an integration region is placed
around the putative source position, with identical background
integration regions distributed around the field of view. The number
of events in the integration region is termed $On$, the number of
events in the background region is termed $Off$ and the ratio of the
integration areas is termed $\alpha$.

\begin{figure}
\begin{center}
\includegraphics [width=0.48\textwidth]{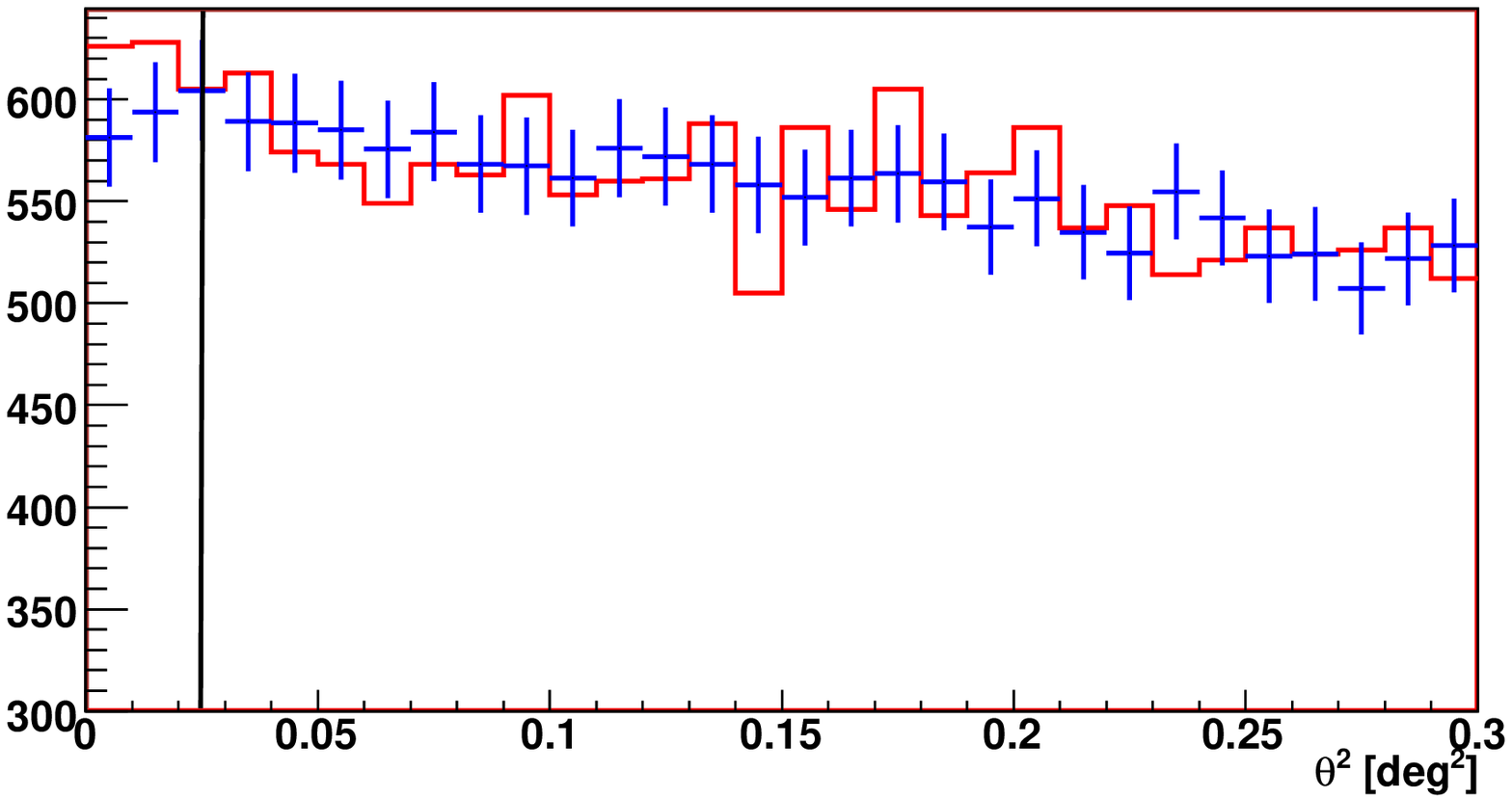}
\end{center}
\caption{$\theta^2$ distribution for 1ES 0647+250 - the signal
integration region is to the left of the vertical
line. The background counts are given by the crosses.}\label{0647_theta}
\end{figure}

\begin{figure}
\begin{center}
\includegraphics [width=0.48\textwidth]{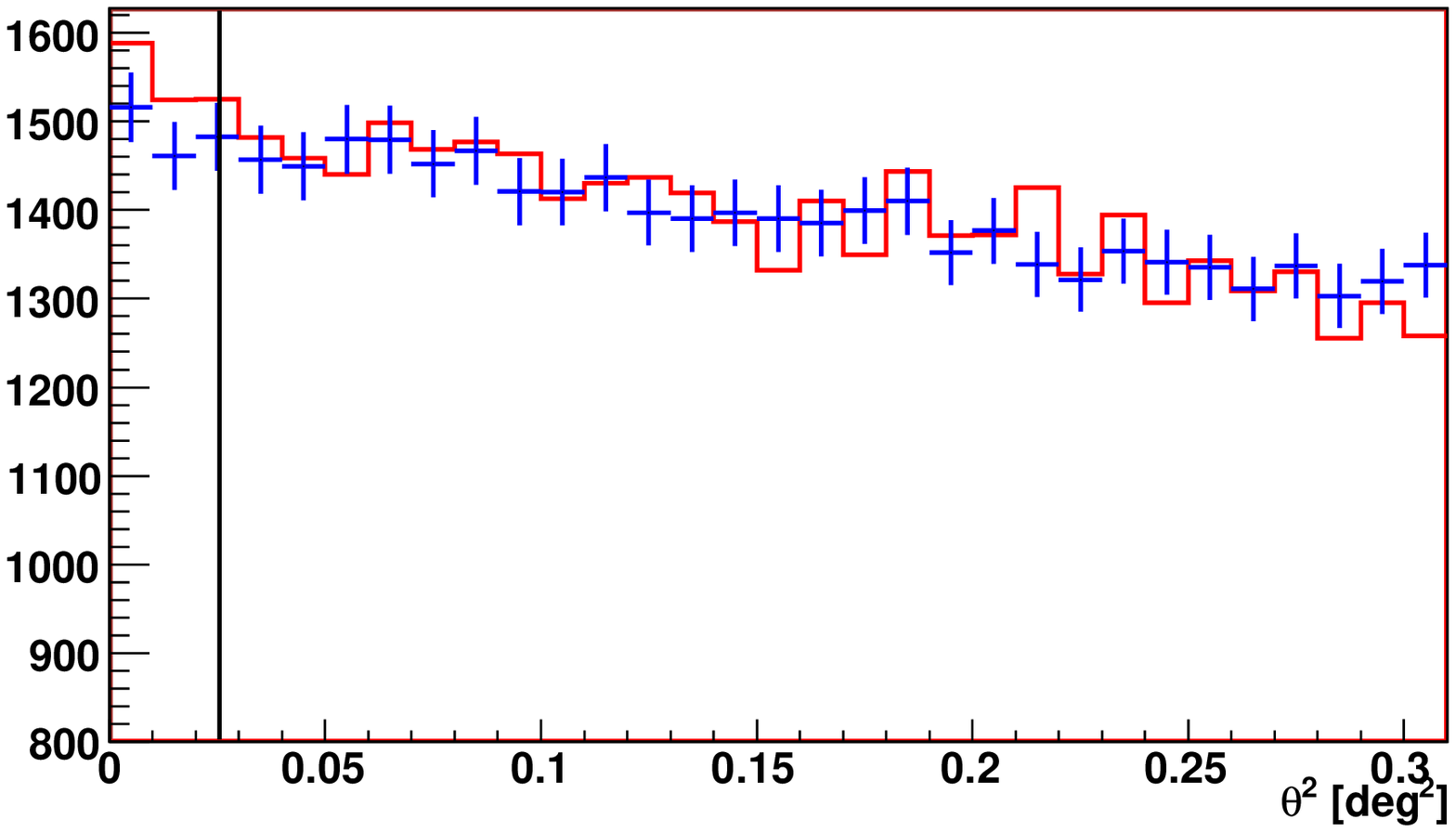}
\end{center}
\caption{$\theta^2$ distribution for 1ES 0806+524 - the signal
integration region is to the left of the vertical
line. The background counts are given by the crosses.}\label{0806_theta}
\end{figure}

The analysis results are summarised in Table \ref{table}, with the
data from both objects broken down in terms of the number of
participating telescopes. The statistical excess is calculated using
equation 17 from \cite{lima83}. No excess above $5\:\sigma$ is
uncovered in any of the data. A distribution of $\theta^2$ for the
signal and background regions for both sources is shown in Figures
\ref{0647_theta} and \ref{0806_theta}. The $\theta^2$ distribution
shows the squared angular distance between the putative source
position and reconstructed shower source. The red line marks the
$\theta^2$ distribution relative to the putative source position and
the blue points mark the averaged $\theta^2$ distribution relative to
the background positions. The signal integration region is to the left
of the vertical line.

In the absence of a clear signal greater than $5\:\sigma$, upper
limits for both objects can be derived. This is done in order to
compare the predictions of the \emph{Costamante} and
\emph{modified-Fossati} models and to previous reports. The upper
limit is found from the probability density function of the number of
counts from the putative source \cite{helene83}.

\begin{equation}
p\: I \left(\frac{-C}{\sigma}\right) = I\left(\frac{C^{\mathrm{UL}}-C}{\sigma}\right)
\end{equation}

where $p$ is the confidence level, $C$ is the number of excess
counts, $\sigma$ is given by

\begin{equation}
\sigma=\sqrt{on + \alpha^2 Off}
\end{equation}

\noindent and the function $I$ is given by

\begin{equation}
I(z)=\frac{1}{\sqrt{2\pi}}\int^\infty_ze^{-x^2/2}\:\mathrm{dx}
\end{equation}

The energy threshold (peak energy response after gamma-ray selection
cuts) for each data set is calculated using a data base of Monte Carlo
gamma-ray simulations at multiple zenith angles. The energy threshold
is found by interpolation in $\cos \Theta$ using the mean
elevation of each data set.

\begin{figure}
\begin{center}
\includegraphics [width=0.48\textwidth]{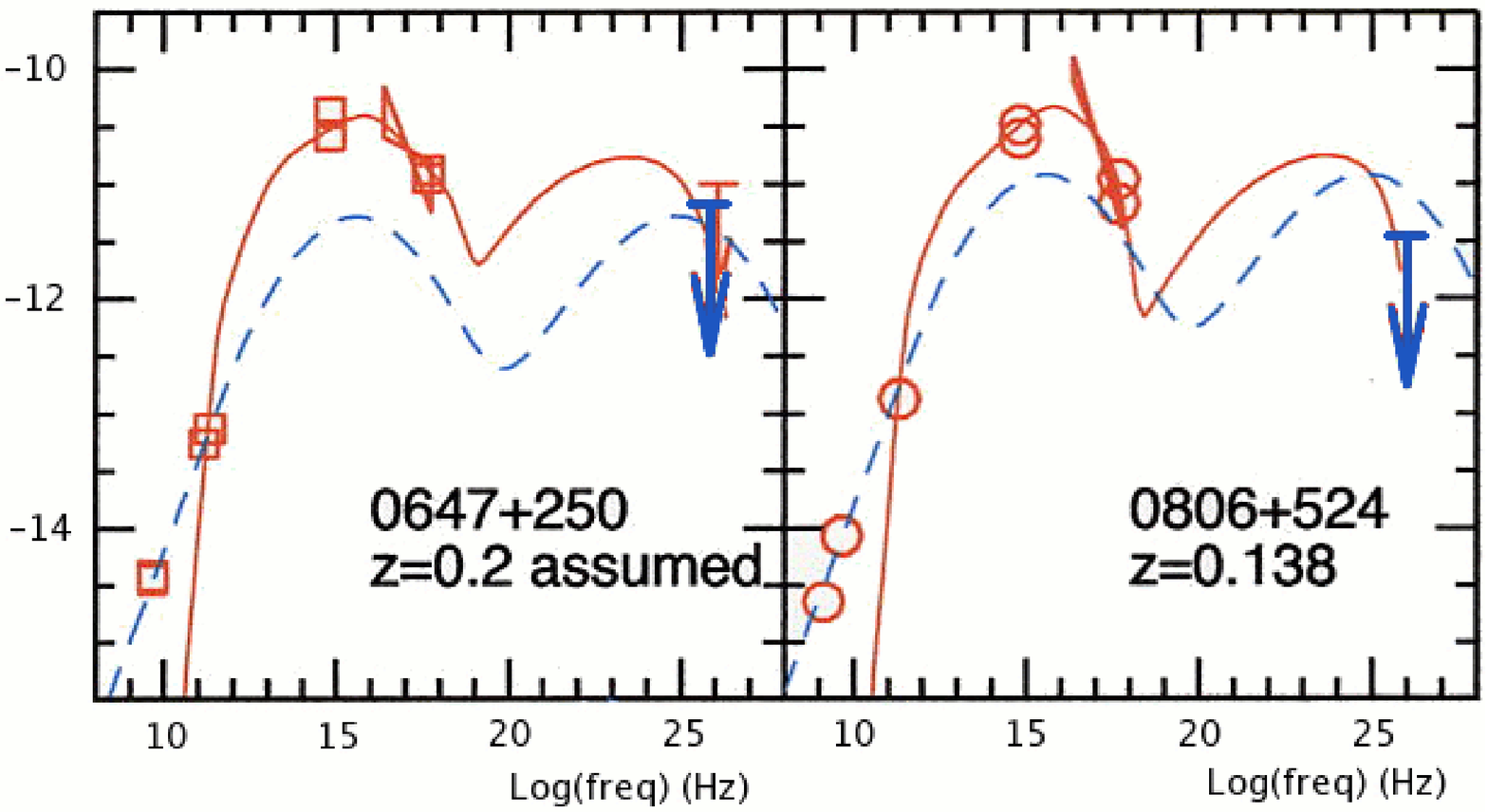}
\end{center}
\caption{Spectral energy distributions of 1ES 0647+250 and 1ES
0806+524 from \cite{costamante02}, with the differential flux upper
limits from this work overlaid in blue. The vertical axis displays
log(vFv) in units of erg cm$^{-2}$ s$^{-1}$.}\label{diff}
\end{figure}

After calculating the $99\:\%$ upper limit on the number of counts. It
is then expressed in terms of Crab Nebula units, and converted into a
flux upper limit by scaling relative to the Crab Nebula flux. These
results are shown in Table \ref{table}. The differential flux upper limits are shown in Figure \ref{diff}.

\begin{table*}[htbp]
  \begin{center}
    \begin{tabular}{ |c|c|c|c|c|c|c|}
      \hline
      Target       & Data Set             & Exposure (hrs) & Excess & $\sigma$  & $\mathrm{E}_{\mathrm{th}}^\mathrm{TeV}$ & $\mathcal{F}^{\mathrm{UL}}_{\mathrm{E}>0.3\:\mathrm{TeV}}$\\ \hline \hline
      1ES 0647+250 & 2-Tel       &  16  & 38 & 1.5 & 0.275 & 0.58\\ \hline 
      1ES 0647+250 & 3-Tel       &  1.3 & 13 & 2.1 & 0.225 & 1.3 \\ \hline 
      1ES 0647+250 & \emph{All}  & 17.3 & 51 & 2.0 & 0.275 & 0.62\\ \hline \hline
      1ES 0806+524 & 2-Tel       &  8.5 & 18 & 1.3 & 0.345 & 0.42\\ \hline 
      1ES 0806+524 & 3-Tel       & 26.2 & 67 & 2.1 & 0.36  & 0.15\\ \hline 
      1ES 0806+524 & \emph{All}  & 34.7 & 85 & 2.5 & 0.36  & 0.13\\ \hline 
  \end{tabular} 
  \end{center}
  \caption{Summary of the VERITAS observations of 1ES 0647+250 and 1ES
  0806+524 from November 2006 to March 2007. The data are categorised
  in terms of the array configuration. The energy threshold,
  $\mathrm{E}_{\mathrm{th}}^\mathrm{TeV}$, is calculated for the average zenith
  angle of the dataset in TeV. The Excess is $\mathrm{\emph{On}} -
  \alpha\mathrm{\emph{Off}}$ and $\sigma$ is the statistical
  significance as calculated using Equation 17 from
  \cite{lima83}. $\mathcal{F}^{\mathrm{UL}}_{\mathrm{E}>0.3\:\mathrm{TeV}}$
  is the flux upper limit above 0.3 TeV in units of
  $10^{-11}\:\mathrm{cm}^{-2}\:\mathrm{s}^{-1}$.}
  \label{table}
\end{table*}

\section{Discussion}

The upper limits calculated for 1ES 0647+250 are significantly better
than those reported previously. The \emph{Costamante} model is surpassed,
however with an extremely low prediction of
$\mathcal{F}_{\mathrm{E}>0.3\:\mathrm{TeV}}=0.24 \times
10^{-11}\:\mathrm{cm}^{-2}\:\mathrm{s}^{-1}$, the
\emph{modified-Fossati} prediction is not surpassed. The upper limits
calculated for 1ES 0806+524 are also significantly better than those
reported before, and surpass the \emph{Costamante} model.

Although predictions for the \emph{Costamante} model on both objects
are surpassed, the implications are not necessarily severe for SSC
modeling, especially given the uncertainties in scaling of flux limits
when the spectrum is unknown. The prediction of TeV fluxes
is extremely sensitive to small variations in model parameters such as
radio and X-ray flux. Also, the flux predictions were built from
multiwavelength data on sources that are known to exhibit
variability. Thus the flux at the lower energy regions of the data
sets may have been different during these VERITAS observations, than
it was when the multiwavelength data from which the model was built.

Given the lack of detectable quiescent emission with a modestly deep
exposure, these objects are unlikely to be the subject of future deep
observations with an instrument of this class. However, given that
optical and x-ray flaring have been linked to detectable increases in
emission in the TeV regime, future campaigns may revolve around
target-of-opportunity triggers from the multiwavelength community.

\section{Conclusions}

No evidence for emission is found from observations of 1ES 0647+250
and 1ES 0806+524. Constraining upper limits were placed on the emission from
1ES 0806+524 in the TeV regime. The VERITAS blazar key science project
promises exciting results in the future as the array reaches maturity.

\section{Acknowledgements}
This research is supported by grants from the U.S. Department of
Energy, the U.S. National Science Foundation, the Smithsonian
Institution, by NSERC in Canada, by Science Foundation Ireland and by
PPARC in the UK.

\bibliography{icrc0365}

\begin{thebibliography}{10}

\bibitem{aharonian04_hegra54}
F.~{Aharonian}.
\newblock {\em AAP}, 421:529--537, 2004.

\bibitem{celick07}
O.~{Celick}.
\newblock {\em These Proceedings}.

\bibitem{cogan07}
P.~{Cogan}.
\newblock {\em These Proceedings}.

\bibitem{costamante02}
L.~{Costamante} and G.~{Ghisellini}.
\newblock {\em AAP}, 384:56--71, 2002.

\bibitem{daniel07}
M.~{Daniel}.
\newblock {\em These Proceedings}.

\bibitem{delacalleperez03}
I.~{de la Calle P{\'e}rez}.
\newblock {\em APJ}, 599:909--917, 2003.

\bibitem{fossati98}
G.~{Fossati}.
\newblock {\em MNRAS}, 299:433--448, 1998.

\bibitem{helene83}
O.~Helene.
\newblock {\em NIM}, 212:319, 1983.

\bibitem{horan04_constraints}
D.~{Horan}.
\newblock {\em APJ}, 603:51--61, 2004.

\bibitem{lima83}
T.-P. {Li} and Y.-Q. {Ma}.
\newblock {\em APJ}, 272:317--324, 1983.

\end{thebibliography}
\bibliographystyle{plain}

\end{document}